\documentclass[12pt]{article}
\input epsf.sty
\usepackage{pdfpages}
\newcommand{\nl}{\hspace{-.65cm}}
\newcommand{\be}{\begin{equation}}
\newcommand{\ee}{\end{equation}}
\newcommand{\ben}{\begin{eqnarray}\displaystyle}
\newcommand{\een}{\end{eqnarray}}

\usepackage{amssymb}
\usepackage{amsmath}
\usepackage{simplewick}
\usepackage[all]{xy}

\def\be{\begin{equation}}
\def\ee{\end{equation}}
\def\ba{\begin{align}}
\def\ea{\end{align}}

 \input{epsf}
 \usepackage{epsfig}

\def\sqr#1#2{{\vcenter{\vbox{\hrule height.#2pt
         \hbox{\vrule width.#2pt height#1pt \kern#1pt
            \vrule width.#2pt}
         \hrule height.#2pt}}}}

\begin{document}

\baselineskip=18pt

\title{\Large{\bf The Black Hole Interior and a Curious Sum Rule}}
\author{Amit Giveon$^{a}$, Nissan Itzhaki$^{b}$ and Jan Troost$^{c}$  }
\date{}
\maketitle
\begin{center}
$^{a}$
Racah Institute of Physics, The Hebrew University, Jerusalem, 91904, Israel\\
$^{b}$ Physics Department, Tel-Aviv University,
Ramat-Aviv, 69978, Israel\\
   $^{c}$Laboratoire de Physique Th\'eorique \\
 Unit\'e Mixte du CRNS et
     de l'\'Ecole Normale Sup\'erieure \\ associ\'ee \`a l'Universit\'e Pierre et
     Marie Curie 6 \\ UMR
     8549 
\\ \'Ecole Normale Sup\'erieure \\
   $24$ Rue Lhomond Paris $75005$, France
\end{center}

 \begin{abstract}
   We analyze the Euclidean geometry
near non-extremal NS5-branes
   in string theory, including regions beyond the horizon and beyond
   the singularity of the black brane. The various regions have an exact
   description in string theory, in terms of  cigar, trumpet
   and negative level minimal model conformal field theories. We
   study the worldsheet elliptic genera of these three superconformal
   theories, and show that their sum vanishes. We speculate on the
   significance of this curious sum rule for black hole physics.
\end{abstract}

\section{Introduction}

The Wick rotation of the Lorentzian black hole geometry to a Euclidean
solution of gravity gives valuable information about quantum black
holes.  For the Schwarzschild solution, the Euclidean geometry reads
\be
ds^2=(1-2M/r)dt^2+\frac{dr^2}{1-2M/r}+r^2d\Omega_2^2 \, .
\label{aaa}
\ee
The region that corresponds to the exterior of the black hole $r > 2M$
is known as the cigar geometry. From the cigar geometry, one 
reads off the relationship between the black hole temperature and mass 
 and the value of the Euclidean action that corresponds to the
entropy of the black hole \cite{Gibbons:1976ue}, and one can fix the Hartle-Hawking wave function \cite{Hartle:1976tp} associated with the Lorentzian extended geometry \cite{Israel:1976ur}.

The region that corresponds to the interior of the black hole, $r<2M$,
does not appear to be physical. In the region between the horizon and
the singularity, $0<r<2M$, the space-time signature is $(-,-,+,+)$. It
includes two time directions ($r$ and $t$).  The big advantage of Wick
rotation is thus lost; instead of getting a Euclidean space in which
the path integral is well defined, we get a divergent path integral.
The region beyond the singularity $r<0$ is again more standard
in that its signature is Euclidean.

In this short note, we study an analogue of the Schwarzschild
geometry which is the geometry of near non-extremal NS5-branes. We consider the three
regions identified above, 
after analytic continuation, and study those regions from
 the perspective of a string worldsheet theory.
We establish a curious sum rule for
the worldsheet elliptic genera associated to the three Euclidean regions.
This could
be viewed as an indication that in string theory the interior of the
black hole plays an important role even after Wick rotation.

\section{The NS5-brane  geometry}

The near horizon Euclidean geometry associated with $k$ near extremal NS5-branes in the
type II superstring is described (at $\alpha'=2$) by the metric and dilaton~\cite{Horowitz:1991cd}:
\begin{eqnarray}
ds^2 &=&  (1-2M/r)dt^2 +  \frac{k}{2 r^2} \frac{dr^2}{1-2M/r}
+  2k d \Omega_3^2 + ds^2_{T^5} \, ,
\nonumber \\
e^{-2 \Phi} &=& \frac{r}{2 k} \, ,
\end{eqnarray}
where $M$ is the energy density above extremality.
There is an $H$-field flux on the three-sphere corresponding to $k$ NS5-branes.
The geometry has a similar causal structure to the Schwarzschild geometry~(\ref{aaa}).

 In a first region I, outside the horizon $r>2M$, we have
a two-dimensional cigar geometry (times an $SU(2)_k$ Wess-Zumino-Witten
model, a five-dimensional flat space, as well as a non-trivial dilaton profile).
It has an exact conformal field theory description in terms of
an $SL(2, \mathbb{R})_k/U(1)$ coset conformal field theory
\cite{Elitzur:1991cb,Mandal:1991tz,Witten:1991yr}.

In a second region II, between the horizon and singularity,
$0<r<2M$, we obtain a bell geometry
with all negative signature, corresponding to an $SU(2)_{-k}/U(1)$
coset conformal field theory at a negative level $-k$.  The fact that the level is negative,
and that we therefore have two time directions is also dictated by the
vanishing of the total central charge of the string theory. The
seeming singularity in the string coupling and metric are absent in the
exact conformal field theory description (see e.g. \cite{Maldacena:2001ky,Kounnas:2007pg} for detailed
discussions).

Finally, region III, the region beyond the singularity,
is obtained by setting $r<0$.
The resulting geometry is that of the trumpet, the vectorially gauged
$SL(2 , \mathbb{R})_k/U(1)$ coset conformal field theory
(see e.g. \cite{Giveon:1994fu} for a review). See figure 1 for a map of the regions
in the Lorentzian black hole in Kruskal coordinates to the conformal field theory target spaces.

\begin{figure}
\centering
\includegraphics[width=0.8\textwidth] {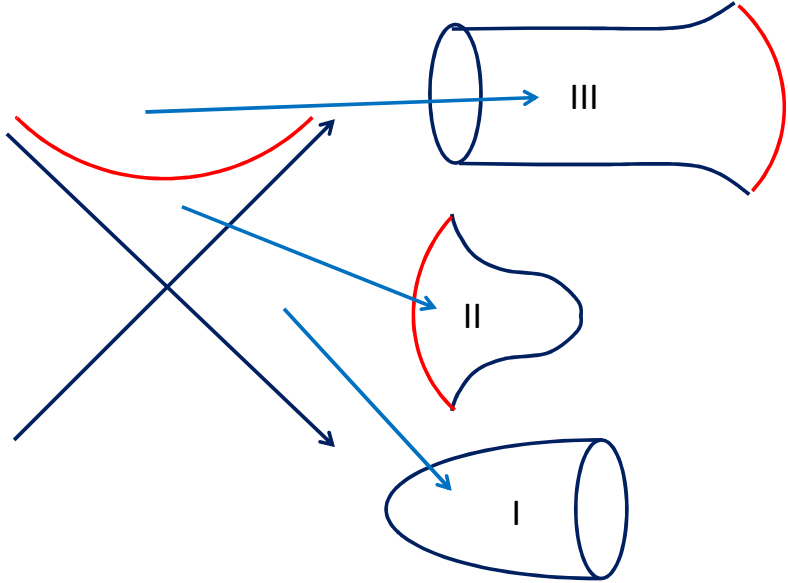}
\caption{The relationship between the Lorentzian black hole in Kruskal coordinates and the different regions obtained after Wick rotation. Regions $I$ and $III$ are Euclidean. Region $II$ includes two time directions. }
\end{figure}

For the regions outside the horizon and beyond the singularity,
 there are exact conformal field theory descriptions. The
elliptic genera of these theories are known
\cite{Troost:2010ud,Eguchi:2010cb,Ashok:2011cy}.
It should be noted that the trumpet
theory is T-dual to the $\mathbb{Z}_k$ orbifold
of the cigar (see e.g. \cite{Giveon:1994fu}).
The region between the horizon and the singularity corresponds
to an $N=2$ minimal
model at a negative level. The  elliptic genus of the conformal field theory with
positive level was calculated longer ago \cite{Witten:1993jg}. We
analytically continue the result  into negative level $-k$ and
we demonstrate below the curious fact that the sum of the three elliptic genera
 vanishes.

\section{The Sum Rule}
The identity
we wish to prove reads:
\begin{eqnarray}
\chi_{I}+\chi_{II}+\chi_{III}\equiv \chi_{cos} (k) +  \chi_{MM} (-k)+\chi_{orb} (k)  &=& 0 \, .
\label{sumrule}
\end{eqnarray}
where the minimal model elliptic genus $\chi_{MM}$ at level $k$ is given by:
\begin{eqnarray}
\chi_{MM}(k) &=& \frac{\theta_{11}(q,z^{1-\frac{1}{k}})}{\theta_{11} (q,z^{\frac{1}{k}})} \, ,
\end{eqnarray}
and the cigar coset and $\mathbb{Z}_k$ orbifold elliptic genera $\chi_{cos} $ and $\chi_{orb}$ are given by:
\begin{eqnarray}
\chi_{cos} &=& \chi_{cos}^{hol} + \chi_{cos}^{rem}
\nonumber \\
\chi_{cos}^{hol} &=& \frac{i \theta_{11}(q,z)}{\eta^3}
\sum_{\gamma=0}^{k-1} \sum_{m \in \mathbb{Z}}
\frac{q^{k m^2- m \gamma} z^{2 m - \frac{\gamma}{k}}}{1-z q^{km -\gamma}}
\nonumber \\
\chi_{cos}^{rem} &=&
- \frac{i \theta_{11} (q,z)}{\pi \eta^3} \sum_{n,w}
\int_{- \infty - i \epsilon}^{+ \infty - i \epsilon}
 \frac{ds}{2is+n+kw} q^{ \frac{s^2}{k} + \frac{(n-kw)^2}{4k}} z^{ \frac{kw-n}{k}}
\bar{q}^{ \frac{s^2}{k} + \frac{(n+kw)^2}{4k}}
\nonumber \\
\chi_{orb} &=& \chi_{orb}^{hol} + \chi_{orb}^{rem}
\nonumber \\
\chi_{orb}^{hol} &=& \frac{i \theta_{11}(q,z)}{\eta^3}
\sum_{m \in \mathbb{Z}}
\frac{q^{k m^2} z^{2 m}}{1-z^{\frac{1}{k}} q^{m}}
\nonumber \\
\chi_{orb}^{rem} &=& -\frac{i \theta_{11} (q,z)}{\pi \eta^3} \sum_{n,w}
\int_{- \infty - i \epsilon}^{+ \infty - i \epsilon}
 \frac{ds}{2is+n+kw} q^{ \frac{s^2}{k} + \frac{(n-kw)^2}{4k}} z^{ \frac{n-kw}{k}}
\bar{q}^{ \frac{s^2}{k} + \frac{(n+kw)^2}{4k}}
\, .
\end{eqnarray}
The proof of the identity proceeds as follows.
In a first step, we notice that the first
two terms in equation (\ref{sumrule}) combine into a holomorphic expression. This is
because of the identity (see the appendix of \cite{Ashok:2012qy}):
\begin{eqnarray}
\chi_{cos}^{rem}
&=&  - \frac{i \theta_{11}}{\eta^3} \sum_{m \in \mathbb{Z}} q^{k m^2} z^{-2 m} - \chi_{orb,rem} \, .
\end{eqnarray}
The second step in the proof uses the theory of elliptic functions, or the theory
of Jacobi forms. In particular, we  follow a proof of the fact that the sum of $N=2$ minimal model characters
and a ratio of theta-functions are equal \cite{DiFrancesco:1993dg}. Thus, we attempt to prove that
the ratio of the two elliptic functions is one:
\begin{eqnarray}
\frac{ \chi_{cos} + \chi_{orb} }{ - \chi_{MM}(-k)} &=& 1 \, .
\end{eqnarray}
We observe that both numerator and denominator have identical modular and elliptic
transformations properties. The ratio is therefore an elliptic function of $z^k$, and a modular
form of weight zero. Moreover both numerator and denominator are holomorphic, due to the reasoning
in step one of the proof. By inspection, it is clear that there are
a finite number of poles and zeroes in the fundamental domain.
It is therefore a ratio of a finite number of theta functions, which equals to one on the condition
that the expression equals one in the $q \rightarrow 0$ limit \cite{DiFrancesco:1993dg}.
This limit is easily taken, and the condition is verified. Therefore, we have proven the identity.

\subsubsection*{Side Remark}
Note that in the special case of level $k=1$ our identity encompasses
a  relation between the conifold and an analytically continued minimal model elliptic genus
\cite{Ooguri:1995wj,Eguchi:2004yi}.  It explains the relative factor of
$\frac{1}{2}$ between the two, which arises from the fact that at level $k=1$, the
cigar and its $\mathbb{Z}_k$ orbifold are identical. Our identity
also correctly captures the overall sign. The latter can be fixed  from
the Witten indices of the individual models.

\section{Interpretations}

A prosaic interpretation of the sum rule is that it is merely
a curious mathematical fact. A second possibility is that this identity
is a consequence of the local physics near the cigar tip, the tip of the
orbifolded cigar, and the T-fold interpretation of the minimal model geometry \cite{Maldacena:2001ky,Kounnas:2007pg}.
An exciting possibility is that the identity
provides us with an important hint about black hole
physics.

Let us entertain the latter possibility. We have a sum
of three different conformal field theory elliptic genera that vanishes. One can view
 the zero on the right hand side of the sum rule (\ref{sumrule}) as the
elliptic genus of a conformal field theory with a vanishing elliptic genus. A natural
guess  for such a conformal field theory is the conformal field theory on the cylinder. Namely the sum rule should be read (see
figure 2):
%
%
\be\label{ty}
\chi_{I}+\chi_{II}+\chi_{III}=\chi(\mbox{cylinder}).
\ee
If so, one
can view the identity as an indication that there is a space associated
to the direct sum of the three conformal field theories associated to the three
Euclidean regions of the black brane geometry, that is topologically a cylinder.
While the two time directions in the region between the horizon and the singularity render
an intuitive interpretation hard, we believe that the point of view in which one takes the
region beyond the horizon seriously even in the Euclidean setting needs to be further explored.

\begin{figure}
\centering
\includegraphics[width=0.8\textwidth] {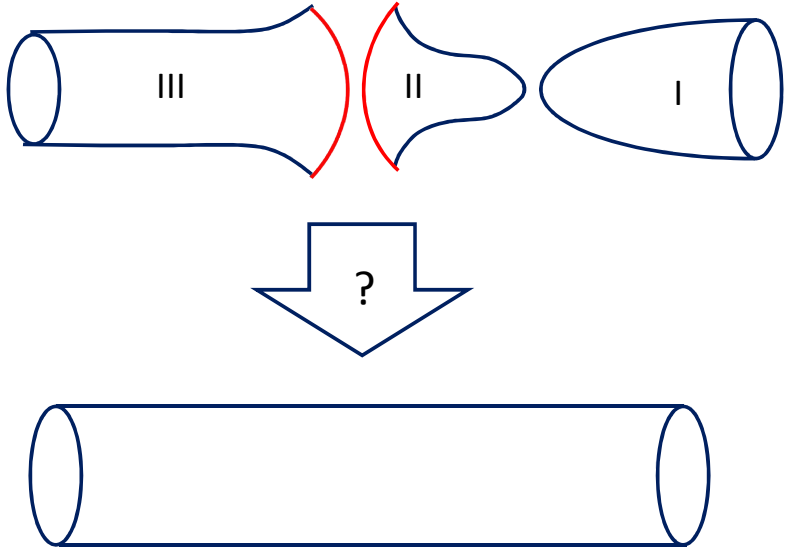}
\caption{Equation (\ref{ty}) suggests that the topology of the 
combined regions is that of a cylinder. }
\end{figure}
\subsubsection*{Note added}
The identity (\ref{sumrule}) was independently found and very
recently published in \cite{Sugawara:2013hma}, where an interpretation
in terms of a compactification of the cigar conformal
field theory was advanced.

\vspace{10mm}

\nl{\bf Acknowledgments}\\
J.T. would like to thank the Hebrew University of Jerusalem as well as Tel-Aviv University for their warm hospitality during his visit.
The work of A.G. and N.I. is supported in part by the I-CORE Program of the Planning and Budgeting Committee
and the Israel Science Foundation (Center No. 1937/12). The work of A.G. is supported in part
by the BSF -- American-Israel Bi-National Science Foundation,
and by a center of excellence supported by the Israel Science Foundation
(grant number 1665/10).

\end{document}